\documentclass[twocolumn,showpacs,preprintnumbers,amsmath,amssymb]{revtex4}
\usepackage[latin1]{inputenc}
\usepackage{amsmath}
\usepackage{amsfonts}
\usepackage{amssymb}
\usepackage[dvips]{graphicx,color}

\begin{document}
\author{C.\ D.\ Fosco}
\email{fosco@cab.cnea.gov.ar}
\affiliation{Centro At\'omico Bariloche,\\
Comisi\'on Nacional de Energ\'\i a At\'omica,\\
8400 S.\ C.\ de Bariloche,\\
Argentina}
\author{J.\ S\'anchez-Guill\'en}
\email{joaquin@fpaxp1.usc.es}
\author{R.\ A.\ V\'azquez}
\email{vazquez@fpaxp1.usc.es}
\affiliation{Departamento de F\'\i sica de Part\'\i culas,\\Facultad de
  F\'\i sica,\\
Universidad de Santiago,\\
E-15706 Santiago de Compostela, Spain}
\title{Tests and applications of  Migdal's particle path-integral 
representation for the Dirac Propagator}

\begin{abstract}
\noindent We derive some non-perturbative results in $1+1$ and $2+1$
dimensions within the context of the particle path-integral representation for
a Dirac field propagator in the presence of an external field, in a
formulation introduced by Migdal in 1986.  
We consider the specific properties of the path-integral
expressions corresponding to the $1+1$ and $2+1$ dimensional cases, presenting
a derivation of the chiral anomaly in the former and of the Chern-Simons
current in the latter.  We also discuss particle propagation in constant
electromagnetic field backgrounds.
\end{abstract}
\maketitle
\section{Introduction}
Particle-like path-integral representations have been used in Quantum
Field Theory since a long time ago, starting with the pioneering work
by Schwinger~\cite{schwinger}.  In this approach, the objects of
interest are expressed in terms of path integrals over particles'
trajectories in proper time, something which is closer in spirit to
Feynman's propagator approach~\cite{feynman} than to standard Quantum
Field Theory methods. Indeed, a propagator is always susceptible of a
dual interpretation: it can be understood as the result of an average
over field configurations, but also as a sum over proper-time
`first-quantized' (i.e., particle-like) paths.

This representation has been more recently applied to the derivation
of many interesting results, since its use provides a framework which
often becomes convenient for the introduction of non-standard
calculation techniques \cite{Schubert}.  
One of the reasons for that, is that the
interaction term appears in an exponential form, and this can make it
possible, sometimes, to integrate out the field that mediates the
interaction.

For the case of non-zero spin fields, different proposals for the
integral over first-quantized trajectories have been advanced. Since
they usually involve different sets of variables, the task of relating
them is far from trivial, unless it is undertaken at a purely formal
level. Concrete calculations, on the other hand, are always useful in
order to understand the properties of each formulation on a deeper
level.

With that in mind, in this article, we consider the particular case of
the path integral representation for Dirac fields introduced by Migdal
in~\cite{migdal}, and apply it as a tool for the derivation of some
non-perturbative results in some Quantum Field Theory models in $1+1$
and $2+1$ dimensions.

This article is organized as follows: in section~\ref{sec:path} we
present a detailed derivation of Migdal's representation, in a way
which is adapted to the applications that we consider afterwards. The
$1+1$ and $2+1$ dimensional cases are discussed in more detail, and a
local action representation suitable for them is introduced. At the
end of this section, the equivalence between the path-integral
representation and the standard formulation is explicitly shown to
be true order by order in perturbation theory.

Going beyond the perturbative expansion, in section~\ref{sec:anom} we
present derivations of the chiral anomaly in $1+1$ dimensions and of
the Chern-Simons term in $2+1$ dimensions; these are two
non-perturbative tests that, as we shall see, reproduce the proper
results.

The propagation in a constant electromagnetic field background in
$2+1$ dimensions is discussed in section~\ref{sec:mag}, by
evaluating the exact fermionic determinant in the present formulation.

Finally, in section~\ref{sec:concl} we present our conclusions.
\section{Path integral representation for the propagator}\label{sec:path}
We shall present here, for the sake of completeness, a derivation of
the particle path-integral representation for the fermion propagator
in an external Abelian gauge field. Besides, the procedure will
emphasize some specific aspects of the $2+1$ and $1+1$ dimensional
cases, as for example the realization of the spin degrees of freedom.
We shall also obtain the standard perturbative expansion within this
framework.
\subsection{Derivation of the general formula}
The propagator for a massive Dirac field in $d$ Euclidean dimensions,
in an Abelian gauge field background, is of course determined by the
(Euclidean) action $S_f$
\begin{equation}\label{eq:desf}
S_f({\bar\psi},\psi,A) \;=\; \int d^dx \, {\bar\psi} (\not\!\! D + m) \psi \;,
\end{equation}
where the $D= -i\partial + e A$ and the $\gamma$-matrices are
Hermitian and verify the Clifford Algebra:
\begin{equation}
\{\gamma_\mu , \gamma_\nu\} = 2 \delta_{\mu\nu} \;.
\end{equation}
$A_\mu$ is an Abelian gauge field, regarded here as external, and $e$ is
a coupling constant with the dimensions of $[{\rm
  mass}]^{\frac{4-d}{2}}$.

The Dirac propagator $G(x,y)$ is the kernel of the inverse of the operator
defining the quadratic form in $S_f$, namely:
\begin{equation}\label{eq:defg}
G(x,y)\;=\; \langle \psi(x) {\bar\psi}(y)\rangle \;=\;
\langle x |( \not \!\!D + m)^{-1} | y \rangle \;,
\end{equation}
where we have adopted Schwinger's convention: $\langle x|K|y\rangle$ for
$K(x,y)$, the kernel of an operator $K$ in coordinate space, and we have
omitted the spinorial indices, although it should be evident from the
context that $\langle x|K|y\rangle$ is a $2\times2$ matrix for $d=2$ and
$d=3$, and a
$4\times4$ matrix when $d=4$.

Assuming $m>0$, we may introduce the exponential representation:
\begin{eqnarray}\label{eq:exprep}
\langle x |( \not \!\!D + m)^{-1} | y \rangle &=& 
 \int_0^\infty dT \; \langle x|U(T)|y\rangle
\end{eqnarray}
defined by the operator:
\begin{equation}\label{eq:defu}
U(T)\;=\; \exp[ - T (\not \!\!D +m) ] \;,
\end{equation}
which acts on functional and spinorial spaces. Note that the
presence of a strictly positive mass $m$ is required for
(\ref{eq:exprep}) to be correct, since we are implicitly assuming the
boundary condition $U(+\infty)=0$, at least in the weak limit
sense.

A functional integral representation can be naturally introduced
to deal with the operator $U(T)$, in spite of the fact that $U(T)$ is
not the exponential of (a constant times) a self-adjoint operator.  As
usual, in a first step, one splits up the `time' $T$ into a number $N$
of intervals of size $\epsilon$, with $T = N \epsilon$. Namely,
\begin{equation}\label{eq:der1}
\langle x|U(T)|y\rangle\;=\;
\langle x|\left\{ \exp[ - \epsilon (\not \!\!D + m) ]\right\}^N|y\rangle \;,
\end{equation}
and then one introduces spectral resolutions of the identity at the
intermediate points $x_1, x_2,\ldots,x_{N-1}$
\begin{eqnarray}\label{eq:der2}
\langle x|U(T)|y\rangle &=& \int(\prod_{k=1}^{N-1} d^dx_k) \;
\langle x|e^{- \epsilon (\not D + m)}|x_{N-1}\rangle \nonumber\\
& & \langle x_{N-1}|e^{- \epsilon (\not D + m)}|x_{N-2}\rangle \ldots
\nonumber\\
&\ldots&\langle x_2|e^{- \epsilon (\not D + m)}|x_1\rangle
\langle x_1|e^{ - \epsilon (\not D + m)}|y\rangle \;.
\end{eqnarray}
It should be kept in mind that the matrix elements on the right hand
side of (\ref{eq:der2}) do not commute with each other; the reason is
of course that each factor is a {\em matrix\/} in spinorial space
rather than a number (as the abbreviated notation might suggest).
Those factors {\em can}, however, be regarded as commuting objects, if they
are put inside a `chronological' ordering symbol ${\mathcal P}$, and
the $\gamma$ matrices are simultaneously given an (auxiliary) dependence on
a discrete time index $k$, that keeps track of their relative positions:
$$\langle x|U(T)|y\rangle \;=\; \int (\prod_{k=1}^{N-1} d^dx_k) {\mathcal
P} \left[ \langle
  x|e^{- \epsilon ( \gamma(N)\cdot D + m ) }|x_{N-1}\rangle \right.$$
$$
\times \langle x_{N-1}| e^{- \epsilon ( \gamma(N-1)\cdot D + m )
}|x_{N-2}\rangle \ldots
$$
\begin{equation}\label{eq:der3}
\left. \ldots  \langle x_2| e^{ - \epsilon ( \gamma(2)\cdot D + m )
}|x_1\rangle \langle x_1|e^{- \epsilon ( \gamma(1)\cdot D + m ) }|y\rangle
\right] \;.
\end{equation}
Then we may write the exact equation:
\begin{eqnarray}\label{eq:der4}
\langle x|U(T)|y\rangle &=&  \int (\prod_{k=1}^{N-1} d^dx_k) \nonumber\\
& {\mathcal P}& \left[ \prod_{l=1}^N \langle x_l|\exp[ -
\epsilon ( \gamma(l)\cdot D + m ) ]|x_{l-1}\rangle \right] \;,
\end{eqnarray}
where we have defined $x_N \equiv x$ and $x_0 \equiv y$.  For each of the
factors under the scope of the ordering operator, we see that:
\begin{eqnarray}\label{eq:der5}
\langle x_l|e^{- \epsilon ( \gamma(l)\cdot D + m ) }|x_{l-1}\rangle &=&
\int d^dp_l \langle x_l|p_l\rangle \nonumber\\
&& \!\!\! \langle p_l | e^{-\epsilon ( \gamma(l)\cdot D + m )} |x_{l-1}\rangle
\end{eqnarray}
which, for $N >> 1$, may be approximated by
\begin{eqnarray}\label{eq:der6}
\langle x_l|e^{- \epsilon ( \gamma(l)\cdot D + m ) }|x_{l-1}\rangle
&\simeq& \int \frac{d^dp_l}{(2\pi)^d}
e^{i p_l \cdot (x_l - x_{l-1})} \nonumber\\
 &\times& \!\!\!\!e^{-\epsilon (i \gamma(l)\cdot(p_l + e A(x_{l-1}) + m ))}
\end{eqnarray}
where we have ignored terms which give no contribution in the $N \to \infty$
limit.

The integration variable $p_l$ is then shifted: $p_l \;\to\; p_l - e
A(x_{l-1})$, with the effect of disentangling the gauge field from the $\gamma$
matrices:
\begin{eqnarray}\label{eq:der7}
\langle x_l|e^{- \epsilon ( \gamma(l)\cdot D + m ) }|x_{l-1}\rangle
&\simeq& \!\! \int
\frac{d^dp_l}{(2\pi)^d} e^{i ( p_l - e A(x_l)) \cdot (x_l - x_{l-1}) }
\nonumber\\
&\times& \!\!  e^{-\epsilon ( i \gamma(l)\cdot p_l + m )} \;.
\end{eqnarray}
Inserting this expression into (\ref{eq:der4}), one sees that:
\begin{eqnarray}\label{eq:der8}
\langle x|U(T)|y\rangle &\simeq& \int (\prod_{k=1}^{N-1} d^dx_k)
(\prod_{l=1}^{N} \frac{d^dp_l}{(2 \pi)^d}) \nonumber\\
& & e^{\epsilon \sum_{l=1}^N[ i p_l\cdot{\dot x}_l - m ]}
{\mathcal P}\left[ e^{-\epsilon \sum_{l=1}^N i \gamma(l)\cdot p_l} \right]
\nonumber\\
&\times&  e^{- i \epsilon e \sum_{l=1}^N {\dot x}_l \cdot A(x_{l-1})}
\end{eqnarray}
where ${\dot x}_l \equiv (x_l - x_{l-1})/\epsilon$.  The time dependence of the
$\gamma$ matrices may now be ignored, since the ordering along the
subdivisions of $T$ is fully determined by the label `$l$' of the $p_l$
which is adjoined to $\gamma_{l}$.

Taking the continuum limit: $N \to \infty$, with $ T = N \epsilon$: fixed,
one obtains the exact (albeit formal) expression:
\begin{eqnarray}\label{eq:cont}
&\langle x|U(T)|y\rangle =  \int {\mathcal D}p {\mathcal D}x \;
e^{ \int_0^T d\tau [ i p \cdot{\dot x} - m ]} &
\nonumber\\
&\times {\mathcal P}[ e^{- i \int_0^T d\tau \, {\not p}}] \,
e^{- i e \int_0^T d\tau {\dot x}(\tau)\cdot A[x(\tau)]} \;,&
\end{eqnarray}
where
\begin{equation}\label{eq:measure}
{\mathcal D}p {\mathcal D}x \;\equiv\; \frac{d^dp(T)}{(2 \pi)^d}\, \prod_{0
< \tau < T}
\frac{d^dx(\tau) d^dp(\tau)}{(2 \pi)^d}\;.
\end{equation}
When used in combination with (\ref{eq:exprep}), (\ref{eq:cont}) yields the
representation for the fermion propagator we were looking for:
\begin{eqnarray}\label{eq:fnal}
&\langle x|(\not\!\!D + m)^{-1} | y \rangle =
\int_0^\infty dT \, \int_{x(0) = y}^{x(T) = x}
{\mathcal D}p {\mathcal D}x &
\nonumber\\
&\times e^{\int_0^T d\tau [ i p\cdot{\dot x} - m ]}
{\mathcal P}[ e^{- i \int_0^T d\tau {\not p}}]
e^{- i e \int_0^T d\tau {\dot x}(\tau)\cdot A[x(\tau)]} \;, &
\end{eqnarray}
where we have indicated explicitly the boundary conditions satisfied
by the paths that have to be integrated out.

It is worth noting the role played by the extra $d^dp$ integration
in the measure, equation (\ref{eq:measure}): each phase-space volume factor
$d^dp d^dx$ is dimensionless, thus the mass dimension of the measure is
determined by the extra $d^dp$ factor. Hence, the
measure has units of $[{\rm mass}]^{d}$. Combining this fact with the
property (self-evident in (\ref{eq:fnal})) that $T$ has dimensions of
$[{\rm mass}]^{-1}$, we see that
the propagator has the dimensions of a $[{\rm mass}]^{d-1}$, as it should be
(twice the mass dimensions of a fermion field).
\subsection{Adiabatic approximation and spin degrees of freedom}
The fact that the functional integral describes the propagation of a spinning
particle manifests itself in the existence of a path-ordered factor
\begin{equation}\label{eq:defphi}
\Phi(T) \;=\; {\mathcal P}\left[ e^{- i \int_0^T d\tau {\not p}
(\tau)}\right] \;,
\end{equation}
whose properties we shall discuss now.

The $d=3$ case is very special, since $\Phi(T)$ allows for a quite
straightforward interpretation as the (quantum) evolution operator for a
spin-$1/2$ in $3$ {\em spatial\/} dimensions, in the presence of a
time-dependent homogeneous `magnetic field' $p_\mu(\tau)$.  Of course,
`evolution' is here understood to mean evolution in the fictitious time
$\tau$. The $3$ components of $p_\mu$ are then regarded as the spatial
components of a magnetic field \mbox{${\mathbf B}=(B_1,B_2,B_3)$}, with
$B_1=p_1$, $B_2=p_2$,$B_3=p_0$.

It should be obvious that, within the crudest infrared approximation
where only constant $p_\mu$ trajectories contribute, $\Phi(T)$ will
not exhibit any interesting behaviour regarding its spin
aspect. Indeed, for a constant magnetic field, one
knows that
\begin{equation}\label{eq:phi1}
\Phi(T) \;=\; e^{- i T p \cdot \gamma  } \;,
\end{equation}
which has the eigenvalues $e^{\mp i T |p|}$, where $|p| \equiv \sqrt{p_\mu
p_\mu}$.

On the other hand, even for a slowly varying $p_\mu$, interesting
effects may, and indeed do arise as a consequence of the existence of
non-integrable Berry's phases, which is a way this representation has
for displaying the non-trivial spin of the field, in the adiabatic
approximation.

For a slowly varying $p_\mu(\tau)$, and assuming $p_\mu(\tau)\neq 0$ to avoid
degeneracy, the adiabatic approximation can be applied to obtain an expression
for $\Phi(T)$. If the initial ($2$-component) state is an eigenstate of
${\not\!p}(0)$, it will, in this approximation, remain an instantaneous
eigenstate during the evolution. At this point, we introduce an explicit
convention for the $d=3$ $\gamma$ matrices: $\gamma_0 = \sigma_3$, $\gamma_1 =
\sigma_1$ and $\gamma_2 = \sigma_2$, where $\sigma_j$, with $j=1,2,3$, denote
the usual Pauli matrices. With this convention, they verify the relation:
\begin{equation}
\gamma_\mu \gamma_\nu \;=\; \delta_{\mu\nu} I
\,+\, i \epsilon_{\mu\nu\lambda} \gamma_\lambda \;.
\end{equation}

Denoting by $|v_\pm(\tau)\rangle$ the instantaneous eigenstates at time
$\tau$, with the eigenvalues $\pm |p(\tau)|$, respectively, we then have the
adiabatic $\Phi(T)$
\begin{eqnarray}
\Phi(T)&\simeq& e^{i \left[ \gamma_+(T) -  \int_0^Td\tau |p(\tau)|
\right]} \,
|v_+(\tau)\rangle\langle v_+(0)|
\nonumber\\
&+& e^{i\left[ \gamma_-(T) +  \int_0^Td\tau |p(\tau)| \right]} \,
 |v_-(\tau)\rangle\langle v_-(0)|
\end{eqnarray}
where $\gamma_{\pm}(T)$ denotes the non-integrable phase corresponding to each
state.

The normalized instantaneous eigenstates $|v_\pm(\tau)\rangle$ can, with
suitable phase conventions, be written as:
\begin{eqnarray}
|v_+(\tau)\rangle &=& \left( \begin{array}{c} \cos\frac{\theta(\tau)}{2}
e^{-i\frac{\phi(\tau)}{2}} \\
\sin\frac{\theta(\tau)}{2} e^{i\frac{\phi(\tau)}{2}}
\end{array}\right)  \; \nonumber\\
|v_-(\tau)\rangle &=& \left( \begin{array}{c} -\sin\frac{\theta(\tau)}{2}
e^{-i\frac{\phi(\tau)}{2}} \\
\cos\frac{\theta(\tau)}{2} e^{i\frac{\phi(\tau)}{2}}
\end{array}\right)
\end{eqnarray}
where $|p|$, $\theta$ and $\phi$ are the spherical coordinates
of the vector $p_\mu$
\begin{eqnarray}
p_0(\tau) &=& |p(\tau)| \cos \theta(\tau) \nonumber\\
p_1(\tau) &=& |p(\tau)| \sin \theta(\tau) \cos \phi(\tau) \nonumber\\
p_2(\tau) &=& |p(\tau)| \sin \theta(\tau) \sin \phi(\tau) \;.
\end{eqnarray}
Then, the spinning nature of the field is evident from the phases
$\gamma_{\pm}(T)$, which are given explicitly by
\begin{eqnarray}\label{eq:defgmpm}
\gamma_{\pm}(T) & \;=\; & i \int_0^T  d\tau \langle v_{\pm}(\tau)|
\frac{d}{d\tau} 
|v_{\pm}(\tau)\rangle \nonumber \\
&\;=\;& \pm \frac{1}{2}
\int_0^T d\tau \,\frac{d\phi(\tau)}{d\tau}  \cos (\theta(\tau)) \;.
\end{eqnarray}
For a closed path ${\mathcal C}$ in the evolution of $p_\mu(\tau)$,
\begin{equation}\label{eq:gmalt}
\gamma_{\pm}(T) \;=\;\pm \frac{1}{2} \int_{\mathcal C}  d\phi \, \cos \theta
\,=\, \pm \frac{1}{2}  \int_{S(\mathcal C)}  d\phi\land d\cos \theta
\end{equation}
where $S({\mathcal C})$ is a regular surface with ${\mathcal C}$ as the
boundary. Note that a closed path appears
when the integral is evaluated with boundary conditions for the
momenta, typically periodic, rather than the coordinates.

It is clear, either in its form (\ref{eq:defgmpm}) or (\ref{eq:gmalt}), that
the phases $\gamma_\pm$ do correspond to actions that can be used for the
quantization of a spin-$1/2$ degree of freedom~\cite{zinn}, in the presence of
an external magnetic field, in $3$ spatial dimensions. The reason for their
appearance here is of course the fact that the Lorentz group in $2+1$
dimensions has been mapped into $SO(3)$ by the Wick rotation. Those groups
have different sets of irreducible representations. The spin-$\frac{1}{2}$
one, however, has a similar meaning and properties for both of them.

It is interesting to compare the situation here with the one in $1+1$
dimensions, where the representation of Dirac's algebra is also constructed in
terms of $2\times2$ matrices, but only two of them appear in
${\not\!p}(\tau)$. It only takes a little amount of thought to see that the
phases $\gamma_\pm(T)$ vanish in this case.

Finally, we comment on the $3+1$ dimensional case. Now the $\gamma$ matrices
in the irreducible representation are of $4 \times 4$ order, however, it is
obvious that the eigenvalues $\Phi(T)$ are still given by the expression
$e^{\mp i T |p|}$, where $|p| \equiv \sqrt{p_\mu p_\mu}$. The main (and
important) difference with the lower dimensional cases is that (being the
$\gamma$ matrices of order $4\times4$), each eigenvalue is doubly degenerated.

Thus the Berry's connection shall be given by a non Abelian $SU(2)$ gauge
field, and as a consequence the expression for the adiabatic phases cannot be
given as explicitly as for the $2+1$ dimensional case.

\subsection{Local action representations in $2+1$ and $1+1$ dimensions}
It may be desirable, in some contexts, to have a path-integral representation
for $\langle x|U(T)|y\rangle$ where the paths are integrated with a local
weight, that can be defined in terms of an action functional.  It is clear
that the factor $\Phi(T)$ is an obstruction to that goal, and that a suitable
local action representation for that object would immediately solve the
problem.

Recalling the magnetic field analogy already used in the previous subsection,
we try to use Grassmann variables to represent the kernel for the $\Phi(T)$
operator.  That this can be done for a spin-$1/2$ particle in a {\em
  constant\/} magnetic field background is a well-known fact. Indeed, for a
Hamiltonian of the form
\begin{equation}
H \;=\; {\mathbf B} \cdot {\mathbf \sigma}
\end{equation}
with a constant ${\mathbf B}$, we may take the $x_3$ axis along the direction
of ${\mathbf B}$, and write $H$ as follows:
\begin{equation}
H \;=\; B (a^\dagger a - a a^\dagger)
\end{equation}
where $a$ and $a^\dagger$ are Fermionic operators: $a^2=0$, $(a^\dagger)^2=0$,
$\{a,a^\dagger\}=1$, and $B=|{\mathbf B}|$. Of course, in this $2$-dimensional
Hilbert space, those operators may be understood as defined by the matrices:
\begin{equation}
a = \left(\begin{array}{cc}0&0\\1&0\end{array}\right)\;\;,\;\;
a^\dagger = \left(\begin{array}{cc}0&1\\0&0\end{array}\right)\;.
\end{equation}
The kernel of the evolution operator can be written in the holomorphic
representation, where the operators defined above act on the space of
`analytic' functions $f(\xi)= a + b \xi$ where $\xi$ is a Grassmann variable,
with $a,b \in {\mathbb C}$, with the scalar product:
\begin{equation}\label{eq:defscprd}
(f,g)\;=\;\int d\xi d{\bar\xi} \, e^{{\bar\xi}\xi} \,
{\overline{f(\xi)}} g(\xi) \;.
\end{equation}
The action of the operators is: $a \to \partial_\xi$, $a^\dagger \to \xi$, and
it is trivial to check that they are adjoint to each other for the scalar
product defined in (\ref{eq:defscprd}).

The kernel of $\exp(-i T H)$ can then be represented as a functional integral,
\begin{equation}\label{eq:holo}
\langle\xi|\exp(-i T H)|{\bar\xi}\rangle \;=\; \int {\mathcal D}\xi
{\mathcal D}{\bar
\xi}\, \exp\left[ - S(\xi,{\bar\xi})\right]
\end{equation}
where
\begin{eqnarray}
S(\xi,{\bar\xi})&=& {\bar\xi}(T) \xi(T) \nonumber\\
&+& \int_0^Td\tau \left[ {\bar\xi}(\tau) {\dot\xi}(\tau)
- 2 i B {\bar\xi}(\tau) \xi(\tau) + i B \right] \;,
\end{eqnarray}
and the paths in the functional integral (\ref{eq:holo}) verify the boundary
conditions $\xi(T) = \xi$ and ${\bar \xi}(0) = {\bar\xi}$.

Things are different when the magnetic field depends on time, since then the
Hamiltonian cannot, in general, be diagonalized by the same similarity
transformation at all times. Indeed, for a general $\tau$-dependent $p_\mu$,
we have to deal with the Hamiltonian:
\begin{eqnarray}\label{eq:hamilt}
H(\tau) &=&  2 p_0(\tau) a^\dagger a - p_0(\tau) + (p_1(\tau) + i
p_2(\tau)) a \nonumber\\
&+& (p_1(\tau) - i p_2(\tau)) a^\dagger \;.
\end{eqnarray}
If the goal is to implement the adiabatic approximation, it is then convenient
to use $H(\tau)$ in terms of its canonical diagonal form:
\begin{equation}
H(\tau) \;=\; |p(\tau)| V^\dagger(\tau) \sigma_3  V(\tau)
\end{equation}
where
\begin{equation}
V(\tau) \;=\; \left(
\begin{array}{cc}
\cos\frac{\theta(\tau)}{2} e^{i\frac{\phi(\tau)}{2}} &
\sin\frac{\theta(\tau)}{2}
e^{-i\frac{\phi(\tau)}{2}} \\
-\sin\frac{\theta(\tau)}{2} e^{i\frac{\phi(\tau)}{2}} &
\cos\frac{\theta(\tau)}{2} e^{-i\frac{\phi(\tau)}{2}}
\end{array}
\right)
\end{equation}
is a unitary matrix that changes the basis to the instantaneous eigenstates.
If now a functional integral representation is introduced, and the adiabatic
approximation is made, it is evident to realize that the evolution operator
will be similar to the one of the case (\ref{eq:holo}), except for the
fact that there will arise a contribution proportional to the diagonal
elements of $\partial_\tau V^\dagger(\tau) V(\tau)$, and these are again the
Berry's phases. Namely, one obtains:
\begin{equation}\label{eq:holotau}
\langle\xi|\Phi(T)|{\bar\xi}\rangle \;=\; \int {\mathcal D}\xi {\mathcal
D}{\bar
\xi}\, \exp\left[ - S_\Phi(\xi,{\bar\xi};T)\right]
\end{equation}
where
$$
S_\Phi(\xi,{\bar\xi};T)\;=\; {\bar\xi}(T) \xi(T) \,+\,
\int_0^Td\tau \left\{ {\bar\xi}(\tau) {\dot\xi}(\tau)
\right.
$$
$$
-  i \;[ p_0(\tau) - \frac{1}{2} \frac{d\phi(\tau)}{d\tau}  \cos
\theta(\tau)\,] \;
{\bar\xi}(\tau) \xi(\tau)
$$
\begin{equation}
\left. +  i \;[ p_0(\tau) - \frac{1}{2} \frac{d\phi(\tau)}{d\tau}  \cos
\theta(\tau)\,]\;
\xi(\tau) {\bar\xi}(\tau) \right\} \;.
\end{equation}

If no approximation is implemented, an exact path-integral representation can
still be written; it corresponds to using an action
\begin{eqnarray}
S_\Phi(\xi,{\bar\xi};T)&=& {\bar\xi}(T) \xi(T)
+ \int_0^Td\tau \left[ {\bar\xi} {\dot\xi}
- i p_0 ( \xi {\bar\xi} - {\bar\xi} \xi ) \right.
\nonumber\\
&-& i \left. (p_1 + i p_2) \xi
-i  (p_1 - i p_2) {\bar\xi}
\right]\;.
\end{eqnarray}
This can be inserted into (\ref{eq:fnal}), to derive the local action
representation:
\begin{eqnarray}\label{eq:fnal1}
\langle x|(\not\!\!D + m)^{-1} | y \rangle &=& \int_0^\infty dT  \int_{x(0) =
  y,{\bar\xi}(0)={\bar\xi}}^{x(T) = x,\xi(T)=\xi} {\mathcal D}p {\mathcal
  D}x {\mathcal D}\xi {\mathcal D}{\bar\xi} \nonumber\\
&\times & \exp[ - {\mathcal S}(p,x,\xi,{\bar\xi};T)]
\end{eqnarray}
where
\begin{eqnarray}
{\mathcal S}(p,x,\xi,{\bar\xi};T)&=& {\bar\xi}(T) \xi(T) + \int_0^Td\tau
\left[- i p\cdot {\dot x} + m \right. \nonumber\\
&+& {\bar\xi} {\dot\xi} -i p_0 ( \xi {\bar\xi} - {\bar\xi} \xi )
- i  (p_1 + i p_2) \xi \nonumber\\
&-& \left. i  (p_1 - i p_2) {\bar\xi}
+ i e {\dot x}\cdot A \right] \;.
\end{eqnarray}

An important remark is in order regarding the last expression. In spite of the
fact that the Grassmannian part of the action looks Gaussian, it cannot be
integrated by the procedure of `completing the square'. Indeed, there is a
difference with the usual Gaussian integral in the fact that the source terms
mix Grassmann and c-number variables. Besides, except for the case when the
adiabatic approximation is used, the local action has the somewhat unpleasant
property of having a non-vanishing Grassmann parity. However, that property is
also present in other formulations of the particle path-integral, since it is
an unavoidable feature of any spinning particle propagator: the fact that it
should be a matrix in some internal space means that we cannot do with a
purely c-number action.

The corresponding result for the propagator in $1+1$ dimensions comes at not
extra price; indeed, adopting the convention that $\gamma_0$ is represented by
$\sigma_1$ and $\gamma_1$ by $\sigma_2$, we see that the analog of
(\ref{eq:hamilt}) is now:
\begin{equation}
H(\tau) \;=\;  (p_0(\tau) + i p_1(\tau)) a + (p_0(\tau) - i p_1(\tau))
a^\dagger \;.
\end{equation}
Thus, in $1+1$ dimensions, we have:
\begin{eqnarray}\label{eq:fnal2}
\langle x|(\not\!\!D + m)^{-1} | y \rangle &=& \int_0^\infty dT  \int_{x(0) =
  y,{\bar\xi}(0)={\bar\xi}}^{x(T) = x,\xi(T)=\xi} {\mathcal D}p {\mathcal
  D}x {\mathcal D}\xi {\mathcal D}{\bar\xi}\nonumber\\
&\times& \exp[ - {\mathcal S}(p,x,\xi,{\bar\xi};T)]
\end{eqnarray}
where
\begin{eqnarray}
{\mathcal S}&=& {\bar\xi}(T) \xi(T)
+ \int_0^Td\tau \left[- i p\cdot {\dot x} + m + {\bar\xi} {\dot\xi} \right.
\nonumber\\
&-& \left. i (p_0+ i p_1)\xi   - i (p_0 - i p_1) {\bar\xi} +
i e {\dot x}\cdot A \right]\;.
\end{eqnarray}

\subsection{Perturbative expansion}
An important check the functional representation must pass, is that it should
reproduce (at least) the perturbative, small-$e$, expansion for the fermion
propagator in an external field.  To do that, we expand the exponential inside
the functional integral of (\ref{eq:fnal}). Using the symbol ${\mathcal
  G}(x,y)$ to denote the functional integral representation (\ref{eq:fnal}),
we see that
\begin{equation}
{\mathcal G}(x,y) \;=\; \sum_{n=0}^\infty {\mathcal G}^{(n)} (x,y)
\end{equation}
where ${\mathcal G}^{(0)}$ is the zeroth-order term,
\begin{eqnarray}\label{eq:ffree}
{\mathcal G}^{(0)} (x,y) &=&\int_0^\infty dT \, \int_{x(0) = y}^{x(T) = x}
{\mathcal D}p {\mathcal D}x \; e^{\int_0^T d\tau [ i p\cdot{\dot x} - m ]}
\nonumber\\
&\times&{\mathcal P}\; \left[ e^{- i \int_0^T d\tau {\not p} (\tau)}\right]\;,
\end{eqnarray}
which, as shown in~\cite{Karanikas:1995zi}, correctly reproduces the
free propagator:
\begin{equation}\label{eq:free}
{\mathcal G}^{(0)} (x,y)\;=\; \int \frac{d^3p}{(2\pi)^3} \,\frac{e^{i p \cdot
(x-y)}}{i\not\! p + m} \;.
\end{equation}
There is an auxiliary identity involving the functional integral that
appears in expression (\ref{eq:ffree}) for the free propagator  that
shall be useful in what follows. It can be derived from the fact
that the integral:
\begin{equation}
\int_{x(0) = y}^{x(T) = x} {\mathcal
D}p {\mathcal D}x e^{\int_0^T d\tau [ i p\cdot{\dot x} - m ]}
{\mathcal P}\left[ e^{- i \int_0^T d\tau {\not p} (\tau)}\right]
\end{equation}
is {\em independent\/} of the boundary values of the momentum. Thus:
\begin{equation}
0=\int_{x(0) = y}^{x(T) = x} {\mathcal D}p {\mathcal D}x
\frac{\delta}{\delta p_\mu(T)} e^{\int_0^T d\tau [ i p\cdot{\dot x} - m ]}
{\mathcal P} e^{- i \int_0^T d\tau {\not p}}
\end{equation}
and an analogous equation for $p(T) \leftrightarrow p(0)$. Then we derive the
identities:
$$
\int_{x(0) = y}^{x(T) = x} {\mathcal D}p {\mathcal D}x \; {\dot x}_\mu(T)
\; e^{\int_0^T d\tau [ i p\cdot{\dot x} - m ]}
{\mathcal P} e^{- i \int_0^T d\tau {\not p} }
$$
\begin{equation}\label{eq:idleft}
= \;  \gamma_\mu \; \int_{x(0) = y}^{x(T) = x} {\mathcal D}p {\mathcal D}x
\; e^{\int_0^T d\tau [ i p\cdot{\dot x} - m ]}
{\mathcal P} e^{- i \int_0^T d\tau {\not p} }
\end{equation}
and
$$
\int_{x(0) = y}^{x(T) = x} {\mathcal D}p {\mathcal D}x \; {\dot x}_\mu(0)
\; e^{\int_0^T d\tau [ i p\cdot{\dot x} - m ]}
{\mathcal P} e^{- i \int_0^T d\tau {\not p} }
$$
\begin{equation}\label{eq:idright}
=\; \int_{x(0) = y}^{x(T) = x} {\mathcal D}p {\mathcal D}x
\; e^{\int_0^T d\tau [ i p\cdot{\dot x} - m ]}
{\mathcal P} e^{- i \int_0^T d\tau {\not p}} \gamma_\mu \;.
\end{equation}
It is important to remember that the functional integral has a matrix-like
weight, so that $\gamma_\mu$ cannot be freely commuted with it.  Besides, in
both of the previous expressions, one cannot move ${\dot x}_\mu(T)$ and ${\dot
  x}_\mu(0)$ out of the integral symbol, since their values are integrated
out, because they are not fixed by the boundary conditions on $x_\mu$.
Finally, both (\ref{eq:idleft}) and (\ref{eq:idright}) can also be easily
proven to hold true in the safer, regulated context of the discretized path
integral.

The term of order $n$ is given by:
\begin{eqnarray}\label{eq:gofn}
{\mathcal G}^{(n)} (x,y) &=& \frac{(-ie)^n}{n!} \int_0^\infty dT
\int_0^Td\tau_1  \ldots \nonumber\\
&\ldots&\int_0^Td\tau_n  \int_{x(0) = y}^{x(T) = x}
{\mathcal D}p {\mathcal D}x \; e^{\int_0^T d\tau [ i p\cdot{\dot x} - m ]}
\nonumber\\
&\times& {\mathcal P} \left[ e^{- i \int_0^T d\tau {\not p} (\tau)}\right]\,
{\dot x}_{\mu_1}(\tau_1) A_{\mu_1}[x(\tau_1)] \ldots \nonumber\\
&\ldots&  {\dot x}_{\mu_n}(\tau_n) A_{\mu_n}[x(\tau_n)]\;.
\end{eqnarray}

The integral over the `intermediate times' $\tau_1,\ldots,\tau_n$ is obviously
symmetric under permutations of the $\tau_i$. For each possible ordering among
them, we now select the maximum time, irrespective of the ordering among the
remaining times. We then rename that maximum time as `$\tau_1$'.  Obviously,
there are $n$ possible contributions to take into account, thus we may write
${\mathcal G}^{(n)}$ in the equivalent way
\begin{eqnarray}
&{\mathcal G}^{(n)} (x,y) = \frac{(-ie)^n}{(n-1)!}  \int_0^\infty dT
\int_0^Td\tau_1 \int_0^{\tau_1}d\tau_2 \ldots \int_0^{\tau_1}d\tau_n &
\nonumber\\
& \times  \int_{x(0) = y}^{x(T) = x} {\mathcal D}p {\mathcal D}x \,
e^{\int_0^T d\tau [ i p\cdot{\dot x} - m ]}
{\mathcal P} \left( e^{- i \int_0^T d\tau {\not p} }\right) &
\nonumber\\
&\times {\dot x}(\tau_1) \cdot A[x(\tau_1)] \ldots
{\dot x}(\tau_n)\cdot A[x(\tau_n)]\;,&
\end{eqnarray}
where the times $\tau_i$, with $i\neq1$, have $\tau_1$ as their new upper
value.

The paths being integrated out in the path integral can then be split at the
time $\tau_1$ by an application of the `superposition principle' for path
integrals, which then requires the value of $x(\tau)$ at the time $\tau_1$ to
be integrated over all its possible values $x(\tau_1) = z$.  Using also
(\ref{eq:idright}), we see that:
\begin{eqnarray}
&{\mathcal G}^{(n)} (x,y) = \frac{(-ie)^n}{(n-1)!}
\int_0^\infty dT  \int_0^Td\tau_1 \int_0^{\tau_1}d\tau_2 \ldots
&\nonumber\\
&\ldots \int_0^{\tau_1}d\tau_n \int d^3z \int_{x(\tau_1)=z}^{x(T) = x}
{\mathcal D}p {\mathcal D}x \, e^{\int_{\tau_1}^T d\tau [ i p\cdot{\dot x}
- m ]}& \nonumber\\
&{\mathcal P} [ e^{- i \int_{\tau_1}^T d\tau {\not p} }] \, {\not \!\!A}(z)
\int_{x(0)=y}^{x(\tau_1)= z} {\mathcal D}p {\mathcal D}x \,
 e^{\int_{0}^{\tau_1} d\tau ( i p\cdot{\dot x} - m )} &\nonumber\\
& {\mathcal P} [e^{- i \int_0^{\tau_1} d\tau {\not p}}]
\, {\dot x}(\tau_2) \cdot A[x(\tau_2)] \ldots
{\dot x}(\tau_n) \cdot  A[x(\tau_n)]\;. &
\end{eqnarray}
Focusing now on the two time integrals which involve the variables $T$ and
$\tau_1$, we see that their order of integration can be interchanged, if one
properly modifies the integration ranges:
\begin{equation}\label{eq:chvar}
\int_0^\infty dT \int_0^{T} d\tau_1 \ldots
\;=\; \int_0^\infty d\tau_1  \int_{\tau_1}^\infty  dT \ldots
\end{equation}
Interchanging those two integrations, and making also some trivial
rearrangements, we see that:
\begin{eqnarray}
&{\mathcal G}^{(n)} (x,y) = -ie \int d^3z  \int_0^\infty d\tau_1
\int_{\tau_1}^\infty dT \, \int_{x(\tau_1)=z}^{x(T) = x}
{\mathcal D}p {\mathcal D}x & \nonumber\\
&\times e^{\int_{\tau_1}^{T} d\tau [ i p\cdot{\dot x} - m ]}{\mathcal P} \left(
  e^{- i \int_{\tau_1}^{T} d\tau {\not p}}\right)\, {\not\!\! A}(z) &
\nonumber\\
&\times \frac{(-ie)^{n-1}}{(n-1)!}  \int_0^{\tau_1}d\tau_2  \ldots
\int_0^{\tau_1}d\tau_n
\int_{x(0)=y}^{x(\tau_1)= z} {\mathcal D}p {\mathcal D}x &\nonumber\\
&\times e^{\int_{0}^{\tau_1} d\tau ( i p\cdot{\dot x} - m )}{\mathcal P} \left(
  e^{- i \int_0^{\tau_1} d\tau {\not p}}\right)& \nonumber\\
&\times  {\dot x}(\tau_2) \cdot A[x(\tau_2)] \ldots
{\dot x}(\tau_n) \cdot A[x(\tau_n)]\;.&
\end{eqnarray}
Noting that
$$
\int_{\tau_1}^\infty dT \, \int_{x(\tau_1)=z}^{x(T) = x} {\mathcal D}p
{\mathcal D}x e^{\int_{\tau_1}^{T} d\tau [ i p\cdot{\dot x} - m ]}{\mathcal P}
\left[ e^{- i \int_{\tau_1}^{T} d\tau {\not p} (\tau)}\right]\,
$$
\begin{equation}
=\; {\mathcal G}^{(0)}(x,z)
\end{equation}
and recalling the expression for the order-$n$ contribution, equation
(\ref{eq:gofn}), we are lead to the relation:
\begin{equation}
{\mathcal G}^{(n)} (x,y) = -  \int d^3z \, {\mathcal G}^{(0)} (x,z)
 \, ie {\not\!\! A}(z) {\mathcal G}^{(n-1)} (z,y)
\end{equation}
($\forall n \geq 1$), which is, indeed, equivalent to the usual perturbative
expansion for the propagator:
\begin{equation}
{\mathcal G}^{(n)}\,=\, {\mathcal G}^{(0)} - ie {{\mathcal G}^{(0)}
  \not \!\! A \mathcal G}^{(0)} + (i e)^2 {{\mathcal G}^{(0)} \not \!\! A
{\mathcal G}^{(0)} \not \!\! A \mathcal G}^{(0)}+\ldots
\end{equation}

It should be evident that the fact that we have not used the `local action'
representation is not crucial to the previous derivations. It is, indeed,
possible to encompass all the changes that proceeding otherwise would produce.
The main differences arise of course in relations (\ref{eq:idleft}) and
(\ref{eq:idright}), since one does not have the $\gamma$ matrices. It is,
however, far from difficult to see that, in the corresponding local action
representation, the equivalent identities relate integrals with components of
${\dot x}_\mu(T)$ to integrals with Grassmann variables. For example:
\begin{eqnarray}
&\int {\mathcal D}p {\mathcal D}x {\mathcal D}\xi {\mathcal D}{\bar\xi}\;
\frac{({\dot x}_1+ i
  {\dot x}_2)}{2} (T) \; \exp[ - {\mathcal S}(p,x,\xi,{\bar\xi};T)]&
\nonumber\\
&= \int {\mathcal D}p {\mathcal D}x {\mathcal D}\xi {\mathcal
D}{\bar\xi}\; {\bar \xi}(T) \; \exp[ - {\mathcal S}(p,x,\xi,{\bar\xi};T)]&\;,
\end{eqnarray}
where both integrals are evaluated with the boundary conditions:
$x(0)=y, \, x(T) = y$ and  ${\bar\xi}(0)={\bar\xi}, \,\xi(T)=\xi $.

On the other hand, integrals involving  ${\dot x}_0$ will be related  to
Grassmann bilinears.
 We shall not proceed, however, the derivation of the perturbative expansion in
that setting, since it would necessarily require, at some point, to
reintroduce the $\gamma$ matrices.

We wish to stress, however, that a perfectly consistent perturbative expansion
could be built in terms of the local representation, without using the
$\gamma$ matrices explicitly.

\section{Chiral anomaly and Chern-Simons current}\label{sec:anom}
In this section we perform another test on the method, with the derivation
of two non-perturbative objects: the chiral anomaly in $1+1$ dimensions and
the Chern-Simons term in $2+1$ dimensions.
They have of course been evaluated in the particle functional integral
framework~\cite{Karanikas:1995ua}; our aim is to show how to obtain them
directly from the functional integral representation(\ref{eq:fnal}), by
evaluating the corresponding vacuum currents. Besides, the role of the
regularization is, as we shall see, more transparent in this calculation.

We shall first deal with the chiral anomaly in $1+1$ dimensions, since this
example already exhibits all the difficulties and properties of the evaluation
of topological terms in this representation. Moreover, we shall use a gauge
invariant
Pauli-Villars regularization, that can be introduced smoothly within the
representation we are dealing with, since it requires the introduction
of (just) one fermion propagator.

It is very well known~\cite{Fujikawa} that ${\mathcal A}$, the
anomalous divergence of the axial current $J_\mu^5 = {\bar\psi}
\gamma_\mu\gamma_5 \psi$
\begin{equation}
\partial_\mu J_\mu^5(x) \;=\; {\mathcal A}(x)
\end{equation}
may be obtained from the regulated trace of the $\gamma_5$ matrix, namely,
\begin{equation}
{\mathcal A}(x) \;=\; \lim_{\Lambda \to \infty} {\mathcal A}_\Lambda (x)
\end{equation}
where
\begin{equation}\label{eq:defal}
{\mathcal A}_\Lambda (x) \;=\; -2 {\rm tr}
\left\{\gamma_5 [ \langle x| f( \frac{i\not \!\! D}{\Lambda})|x\rangle ]
\right\}
\end{equation}
where $f$ is a function chosen in order to tame the UV divergences, what means
that it has to satisfy:
\begin{equation}
f(0) = 1 \;,\;\;f(\pm \infty) = f'(\pm \infty) = f^{(2)}(\pm \infty) =
\ldots  = 0 \;.
\end{equation}
Of course, this is a gauge-invariant regularization, and moreover the results
are independent of the detailed form of $f$, as long as it verifies the
previous conditions~\cite{Fujikawa}.  The particular choice:
$f(u) = {(1+u^2)}^{-1}$ is very convenient, since we can use a simple
fractions decomposition and the fact that $\gamma_5$ anticommutes with
$\gamma_\mu$ to write (\ref{eq:defal}) as follows:
\begin{equation}\label{eq:anom1}
{\mathcal A}_\Lambda (x) \;=\;
- 2 {\rm tr} \left\{\gamma_5 [ \langle x|\frac{\Lambda}{ \not\!\!
    D + \Lambda}|x\rangle ] \right\} \;.
\end{equation}
Then we apply the general expression (\ref{eq:fnal}), with $x=y$ and $m =
\Lambda$, to write the fermion propagator that appears in (\ref{eq:anom1}) as
a particle path-integral, obtaining
$${\mathcal A}_\Lambda (x) \;=\;
- 2  \Lambda \int_0^\infty dT \, \int_{x(0) = x}^{x(T) = x}
{\mathcal D}p {\mathcal D}x \, e^{\int_0^T d\tau [ i p\cdot{\dot x} -
\Lambda  ]} $$
\begin{equation}\label{eq:anom2}
 \times {\rm tr} \left\{ \gamma_5
{\mathcal P} [e^{- i \int_0^T d\tau {\not p} (\tau)}]
\right\}
e^{- i e \int_0^T d\tau {\dot x}(\tau)\cdot A[x(\tau)]} \;.
\end{equation}
The constant $\Lambda$ may be absorbed in a redefinition of $T$: $t = \Lambda
T$ is now a dimensionless `time', while we also introduce $s = \Lambda \tau$
for the `proper time' that appears inside the integrals. Then, expressing all
the functions in terms of the new variables,
$${\mathcal A}_\Lambda (x) \;=\; - 2 \int_0^\infty dt \, \int_{x(0) =
x}^{x(t) = x}
{\mathcal D}p {\mathcal D}x \, e^{\int_0^t ds [ i p\cdot{\frac{dx}{ds}} - 1
]} $$
\begin{equation}\label{eq:anom3}
 \times {\rm tr} \left\{ \gamma_5
{\mathcal P} [e^{- \frac{i}{\Lambda} \int_0^t ds {\not p} (s)}]
\right\}
e^{- i e \int_0^t ds {\dot x}(s)\cdot A[x(s)]} \;,
\end{equation}
where the $x$ which appears in the boundary conditions does not have
to be integrated, but it is the (fixed) value corresponding to the
argument of the current operator.

Following a similar technique (but a different notation) to the one used
in~\cite{Karanikas:1995ua} to evaluate the $\theta$ vacua term, we introduce
the change of variables:
\begin{equation}
x_\mu(s) \;=\; x_\mu(0) \,+\, \int_0^s d{\tilde s} \zeta_\mu ({\tilde s})
\;\;,
\;\; x(0) \equiv x \;,
\end{equation}
which has a trivial Jacobian, so that ${\mathcal D}x = {\mathcal D}\zeta$, and
(\ref{eq:anom3}) becomes
$${\mathcal A}_\Lambda (x) \;=\; - 2 \int_0^\infty dt \, \int{\mathcal D}p
{\mathcal D}\zeta \, e^{\int_0^t ds [ i p\cdot\zeta - 1 ]} $$
\begin{equation}\label{eq:anom4}
 \times {\rm tr} \left\{ \gamma_5 {\mathcal P} [e^{- \frac{i}{\Lambda}
\int_0^t ds {\not p} (s)}]
\right\}
e^{- i e \int_0^t ds \zeta(s)\cdot A[x(0)+\int_0^s d{\tilde s}
\zeta({\tilde s}) ]} \;.
\end{equation}
It is important to note that the boundary conditions for $x$ in
(\ref{eq:anom2}) mean that the $\zeta$ variable has to verify the constraints:
\begin{equation}
\int_0^t ds \zeta_\mu (s) \;=\; 0 \;\;, \; \mu = 0, 1\;;
\end{equation}
the existence of those constraints will be indicated by an $'$ in the
integral symbol, when rewriting the $x$-integral of (\ref{eq:anom3})
in terms of the new variables.

Then we take advantage of the fact that the result has to be a local
polynomial in $A$, to write:
$${\mathcal A}_\Lambda (x) \;=\; - 2 \int_0^\infty dt \, \int {\mathcal D}p
{\rm tr}
\left\{ \gamma_5 {\mathcal P} [e^{- \frac{i}{\Lambda} \int_0^t ds {\not p}
(s)}]\right\}
$$
\begin{equation}\label{eq:anom5}
e^{- i e \int_0^t ds \frac{\delta}{i \delta p(s)}\cdot A[x(0)+\int_0^s
d{\tilde s}
  \frac{\delta}{i \delta p({\tilde s})} ]}
\int' {\mathcal D}\zeta  \, e^{\int_0^t ds [ i p\cdot\zeta  - 1 ]} \;.
\end{equation}
The functional integral over $\zeta$, including the constraint, can be
explicitly evaluated, for example by including the constraint through
the addition of (yet) another Lagrange multiplier, i.e.,
$$
\int'{\mathcal D}\zeta  \, e^{\int_0^t ds [ i p\cdot\zeta  - 1 ]}
\;=\;\int {\mathcal D}\zeta \int
\frac{d^2w}{(2\pi)^2} \, e^{i \int_0^t ds ( p\cdot\zeta  + w
  \cdot\zeta ) } e^{-t}
$$
\begin{equation}\label{eq:intzeta}
=\; \int \frac{d^2w}{(2\pi)^2} \, \delta[p + w] e^{-t}\;.
\end{equation}
It should be noted that $\delta[p - w]$ is a functional $\delta$, and that the
integral over $w$ is a relic of the constraint over the $\zeta$ integration.
Thus we arrive to a more tractable expression for the anomaly
\begin{eqnarray} \label{eq:anom6}
&{\mathcal A}_\Lambda (x)=-  2 \int_0^\infty dt  e^{- t}
\int\frac{d^2w}{(2\pi)^2} \int {\mathcal D}p
{\rm tr}( \gamma_5 {\mathcal P} e^{- \frac{i}{\Lambda}
\int_0^t ds {\not p}}) &\nonumber\\
&e^{- i e \int_0^t ds \frac{\delta}{i \delta p(s)}\cdot A[x(0)+\int_0^s
d{\tilde s} \frac{\delta}{i \delta p({\tilde s})} ]} \delta[p + w]&
\end{eqnarray}
A simple power-counting argument shows that by expanding the $e$ dependent
term in the exponential, only the first order term will contribute when
\mbox{$\Lambda \to \infty$}. Moreover, in that term just the one which is of
order 2 in the functional derivative over $p$ survives (as it is seen
{\em a posteriori}).  Thus, the $p$ integration can also be explicitly
performed:
\begin{equation}
{\mathcal A}_\Lambda (x) \;\simeq\; \partial_\mu A_\nu(x) \; T_{\mu\nu}
\;\;(\Lambda \sim \infty)
\end{equation}
where $T_{\mu\nu}$ is a constant tensor, given explicitly by:
\begin{equation}\label{eq:tmyny}
T_{\mu\nu}= - 2 i e \int\frac{d^2w}{(2\pi)^2} \int_0^\infty dt
e^{- t} \int_0^t ds \int_0^s ds'
\frac{\delta^2 \Phi[p]}{\delta p_\nu(s) \delta p_\mu(s')}|_{p = -\omega}
\end{equation}
where
\begin{equation}
\Phi[p] \;=\;{\rm tr} \left[\gamma_5{\mathcal P} e^{-\frac{i}{\Lambda}
\int_0^t ds {\not p}(s)}\right]\;,
\end{equation}
and $p$ in (\ref{eq:tmyny}) is set equal to the constant $-\omega$ {\em
after\/} the functional differentiation.

For the evaluation of $T_{\mu\nu}$, we note that
\begin{eqnarray}
&T_{\mu\nu}\;=\; i \frac{e}{\Lambda^2} \int\frac{d^2w}{(2\pi)^2}
\int_0^\infty dt t^2 
e^{- t} {\rm tr} \left[\gamma_5 \gamma_\mu \gamma_\nu e^{
    i \frac{t}{\Lambda} \, \not  \omega} \right]&\nonumber\\
& = -  \epsilon_{\mu\nu} \;  \frac{e}{\Lambda^2}
\int\frac{d^2w}{(2\pi)^2} \int_0^\infty dt\,
t^2 \, e^{- t} {\rm tr} \left[e^{i \frac{t}{\Lambda}
\not \omega } \right]\; ,&
\end{eqnarray}
where we have used that
\begin{equation}
\gamma_\mu \gamma_\nu = \delta_{\mu \nu } I + i \epsilon_{\mu \nu} \gamma_5,
\end{equation}
with $I$ the identity matrix. Finally,
\begin{equation}
T_{\mu\nu}\;=\; -  \epsilon_{\mu\nu} \;  \frac{2 e}{\Lambda^2}
\int\frac{d^2w}{(2\pi)^2}
{\rm tr} \left[\frac{1}{(1 - \frac{i}{\Lambda}\not \! \omega)^3}\right]\;.
\end{equation}
The $\omega$ integration is convergent, and its result is proportional to
$\Lambda^2$, what cancels out the $\Lambda^{-2}$ factor.
\begin{eqnarray}
T_{\mu\nu} &=& -  \epsilon_{\mu\nu} \;  \frac{e}{\pi}  \int_0^\infty dx
\frac{ 1 - 3 x }{(1 + x)^3} \nonumber\\
&=&  \frac{e}{\pi} \; \epsilon_{\mu\nu} \;.
\end{eqnarray}
Then we conclude that the anomaly ${\mathcal A}(x)$ has the proper result,
namely,
\begin{equation}
{\mathcal A}(x) \;=\;  \frac{e}{\pi} \epsilon_{\mu\nu} \partial_\mu A_\nu(x)\;.
\end{equation}
Note that, because of the gauge invariant regularization procedure, we already
knew that $T_{\mu\nu}$ had to be proportional to $\epsilon_{\mu\nu}$.

Let us conclude this section with the evaluation of a related object: the
vacuum Chern-Simons current for the Abelian case, in $2+1$ dimensions, as
determined by the parity anomaly.  The vacuum Chern-Simons current, in a
Pauli-Villars like regularization, is given by
\begin{equation}
J_\mu(x) \;=\; - {\rm tr}\left[\gamma_\mu \langle x| (\not \!\! D + M
)^{-1}|x\rangle \right]
\end{equation}
where $M \to \infty$, and all the objects are assumed to be defined in $2+1$
Euclidean dimensions. An entirely analogous derivation yields,
\begin{equation}
J_\mu(x) \;\simeq \; - \; R_{\mu\nu\lambda} \partial_\nu A_\lambda  \;\;
(\Lambda \sim \infty)
\end{equation}
where
\begin{eqnarray}
R_{\mu\nu\lambda}&=& - i \frac{e}{\Lambda} \int\frac{d^3w}{(2\pi)^3}
\int_0^\infty dt  e^{- t} \nonumber\\
&\times&\int_0^s ds'\frac{\delta^2 \Gamma_\mu[p]}{\delta p_\lambda(s)
\delta p_\nu(s')}|_{p = \omega.}
\end{eqnarray}
with
\begin{equation}
 \Gamma_\mu[p] \;=\; {\rm tr} \left[\gamma_\mu{\mathcal P}
e^{-\frac{i}{\Lambda} \int_0^t ds {\not p}(s)}\right]\;.
\end{equation}

More explicitly:
\begin{equation}
R_{\mu\nu\lambda}\;=\; - \frac{e}{\Lambda^3} \, \epsilon_{\mu\nu\lambda} \,
\int\frac{d^3w}{(2\pi)^3} {\rm tr}\left[\frac{1}{(1 - i \not
    \!\omega)^3} \right]\;.
\end{equation}
A standard evaluation of the momentum integral yields
\begin{equation}
R_{\mu\nu\lambda}\;=\;   \frac{e}{2 \pi } \, \epsilon_{\mu\nu\lambda} \;.
\end{equation}
Finally,
\begin{equation}
J_\mu \;=\; \frac{e}{2 \pi} \; \epsilon_{\mu\nu\lambda} \partial_\nu
A_\lambda \;,
\end{equation}
which again reproduces the correct result.

It is worth noting that no extra regularization has been introduced in order
to regularize the anomaly and the Chern-Simons term, what is a difference with
previous calculations in this context. Some of them seem to indicate that a
space resolution scale plays an important role. We have seen, however, that
the usual UV regulator does the trick.

\section{Propagation in a constant electromagnetic field}\label{sec:mag}
We calculate here the fermionic determinant for a massive Dirac field
in the presence of an external constant electromagnetic field, in
$2+1$ Euclidean dimensions. We define,
\begin{equation}\label{eq:defga}
\exp[ - \Gamma(A) ] \;=\; \det (\not \!\! D + m )
\end{equation}
where the gauge field entering the covariant derivative is such that:
\begin{equation}
\partial_\mu A_\nu - \partial_\nu A_\mu \;=\; F_{\mu\nu} = {\rm constant} \;.
\end{equation}
In a symmetric or coordinate gauge, we can adopt the configuration
\begin{equation}\label{eq:aoff}
A_\mu(x) \;=\; - \frac{1}{2} F_{\mu\nu} x_\nu \;=\; \frac{1}{2} 
\epsilon_{\mu\rho\nu}
{\tilde F}_\rho \, x_\nu \;.
\end{equation}
We consider here a gauge-field configuration where only two components
of $A_\mu$ are non-vanishing, and lead to a constant electromagnetic
field. The parity-odd part of the effective action is zero for this sort
of configuration, since:
\begin{equation}
A_\mu \epsilon_{\mu\nu\rho} \partial_\nu A_\rho = A_\mu {\tilde F}_\mu = 0.
\end{equation}

To evaluate $\Gamma(A)$, we  use an integral representation for the
logarithm; as usual we apply the formula
\begin{equation}\label{eq:frull}
\ln (a/b) \;=\; - \int_{0}^\infty \frac{dT}{T} \, \left(e^{- T a} - e^{-
T b} \right)
\end{equation}
($\Re(a) > 0$, $\Re(b)>0$) so that:
\begin{eqnarray}\label{eq:defgta}
{\tilde \Gamma}(A) &\equiv& \Gamma(A) - \Gamma(0) \nonumber\\
&=&  \int_{0}^\infty \frac{dT}{T} \,
{\rm Tr} \left[ e^{- T (\not  D + m) } - e^{- T (\not  \partial + m)}
\right] \;.
\end{eqnarray}
Introducing then the particle path-integral representation, we see that
\begin{eqnarray} \label{eq:gppa}
& \Gamma (A) = \int_{0} \frac{dT}{T} \, e^{- m T} \, \int {\mathcal D}p
\, {\rm tr}\left( {\mathcal P} e^{- i \int_0^T d\tau \not p }
\right) & \nonumber\\
&\times \int {\mathcal D} x \; e^{\frac{i e}{2} \int_0^T d\tau  {\dot
    x}_\mu F_{\mu\nu} x_\nu + i \int_0^T d\tau p_\mu
{\dot x}_\mu }\;, &
\end{eqnarray}
which differs from (\ref{eq:defgta}) by an (infinite) field-independent
constant.

We now proceed to evaluate $I[p]$, the Gaussian integral over
periodic $x_\mu(\tau)$ paths in (\ref{eq:gppa}). After
an integration by parts
\begin{eqnarray}\label{eq:gauss}
& I[p]=\int_{x(0) = x(T)}  {\mathcal D} x \; \exp \left[\frac{i e}{2} \int_0^T
  d\tau \; x_\mu(\tau) \epsilon_{\mu\rho\nu} {\tilde F}_\rho {\dot
    x}_\nu(\tau) \right. & \nonumber\\
& \left. - i \int_0^T d\tau x(\tau) \cdot {\dot p}(\tau) \,+\, 
x(0) \cdot (p(T) - p(0) ) \right] \;.&
\end{eqnarray}
The role of the last term is, after integrating over $x(0)$, to enforce the
condition $p(0) = p(T)$. We can then erase that term when evaluating $I[p]$,
while keeping in mind the fact that the remaining functional integral (over
$p$) in (\ref{eq:gppa}) must the be evaluated with periodic boundary
conditions
for $p(\tau)$.

We now decompose $x_\mu(\tau)$ into parallel ($x^\parallel$) and transverse
($x^\perp$) components along ${\tilde F}$,
\begin{equation}
x_\mu(\tau) \;=\; x^\parallel_\mu(\tau) \,+\, x^\perp_\mu(\tau)
\end{equation}
where $x^\parallel_\mu(\tau)$ is the projection of $x_\mu(\tau)$ along the
direction of ${\tilde F}_\mu$, and $x^\perp_\mu(\tau)$ its orthogonal
component ($x^\perp \cdot {\tilde F} = 0$).  It is evident that the parallel
component does not appear in the quadratic part of the exponent in
(\ref{eq:gauss}), so that its integration yields a functional $\delta$ of the
derivative of the longitudinal component of the momentum:
\begin{eqnarray}\label{eq:gauss1}
I[p]&=&\delta[{\dot p}^\parallel] \frac{L}{2\pi} 
\; \int_{x^\perp(0) = x^\perp(T)} \,
{\mathcal D} x^\perp  \nonumber\\
&\times& e^{\frac{i e}{2} \int_0^T d\tau
  x^\perp_\mu \epsilon_{\mu\rho\nu} {\tilde F}_\rho  {\dot x}^\perp_\nu
- i \int_0^T d\tau \, x^\perp_\mu {\dot p}^\perp_\mu } \;,
\end{eqnarray}
where $L$ is the (infinite) size of the integration in the parallel direction.
The Gaussian integral over the two remaining components of $x$ can now
be performed, and the result may be written in terms of the two
transverse components of the momentum. Using a coordinate system such
that $p_1$ and $p_2$ denote the transverse components, while ${\tilde
  F}_\mu$ points in the $0$ direction, we see that:
\begin{eqnarray}
I (p) &=& \delta[{\dot p}^\parallel] \frac{L}{2\pi} \; 
\int {\mathcal D} x_1 {\mathcal D} x_2 \;
\exp [ i e {\tilde F}_0 \nonumber \int_0^T d\tau {\dot x}_1  x_2\\ 
&+ & 
 i \int_0^T d \tau ( {\dot x}_1(\tau) p_1 (\tau)
                      + {\dot x}_2(\tau) p_2 (\tau) ) ].
\end{eqnarray}
Introducing now an auxiliary variable $w_1$ and the velocity
$\zeta_1$ for the $x_1$ coordinate we have 
\begin{eqnarray}
I (p) & = & \delta[{\dot p}^\parallel] \frac{L}{2\pi} \; 
\int \frac{dw_1}{2 \pi} {\mathcal D} \zeta_1 {\mathcal D}x_2
\; \exp [ i e {\tilde F}_0 \int_0^T d\tau \zeta_1(\tau) x_2(\tau)
  \nonumber \\
& + & i \int_0^T d \tau ( \zeta_1(\tau) (p_1 (\tau) + w_1)
                      + {\dot x}_2(\tau) p_2 (\tau)  ) ],
\end{eqnarray}
where $w_1$ is independent of $\tau$. Integrating now over 
$x_2$ and $\zeta_1$  we arrive to 
\begin{eqnarray}
I (p) & = & \delta[{\dot p}^\parallel] \frac{L}{2\pi} \; 
 \int \frac{dw_1}{2 \pi} \; \int dx_2(0) \nonumber \\
&\times& \exp [ \frac{i }{ e {\tilde F}_0}  
\int_0^T d\tau  {\dot p}_1 (\tau) p_2
(\tau) \nonumber \\
&+ & i \frac{1}{e {\tilde F}_0} w_1 (p_2(T) - p_2(0)) ]
\end{eqnarray}
The integral over $x_2(0)$ is proportional to the
total length of the space. On the other hand the integral 
over $w_1$ gives a delta function which produces a $e {\tilde F}_0$
factor, and an additional total length factor. So we arrive at
\begin{eqnarray}
I[p] &= & \delta[{\dot p}^\parallel] \frac{V}{(2\pi)^2} e {\tilde F}_0 \;
[\det (\frac{ - e {\tilde F}_0 \epsilon_{jk} \partial_\tau }{2 \pi i})]^{-
  \frac{1}{2}} \nonumber \\
& \times & e^{\frac{i}{2 e {\tilde F}_0} \, \int_0^T d\tau \,
p_j \epsilon_{jk} \partial_\tau  p_k} \;,
\label{eq:gauss2}
\end{eqnarray}
where $V$ is the total (infinite) space time volume.
For the sake of simplicity, we shall assume in what follows that $e {\tilde
  F}_0 >0$.

Using the result (\ref{eq:gauss2}), we may now calculate the remaining
functional integral over $p_\mu$ in (\ref{eq:gppa}). Indeed, the longitudinal
component of $p$ shall be constant, so it can be extracted out of the
path-ordering symbol, while for the two transverse components we take
advantage of the following fact: functional integrating an ordered product of
transverse components with the quadratic weight above amounts to taking a
trace over the Hilbert space corresponding to the two non commuting operators
${\hat p}_1$ and ${\hat p}_2$:
\begin{equation}
[{\hat p}_1 , {\hat p}_2 ]\;=\; - i e {\tilde F}_0 \;.
\end{equation}
More explicitly, we repeatedly use the property:
\begin{eqnarray}
&\frac{\int_{p(0)=p(T)} {\mathcal D}p^\perp \,{\mathcal P}[p_{i_1}(\tau_1)
  \ldots p_{i_n}(\tau_n)] e^{ \frac{i}{2 e {\tilde F}_0} \, \int_0^T
    d\tau p_j \epsilon_{jk} \partial_\tau p_k}}{\int
  {\mathcal D}p^\perp \; e^{ \frac{i}{2 e {\tilde F}_0} \, \int_0^T
    d\tau p_j \epsilon_{jk} \partial_\tau p_k}} &
\nonumber\\
&= {\rm Tr} \left[ {\hat p}_{i_1}(\tau_1) \ldots {\hat p}_{i_n}(\tau_n)
\right] \;, &
\label{eq:gauss3}
\end{eqnarray}
where the indices $i_1 \ldots i_n$ can only take the values $1$ and $2$. A
possible way to prove (\ref{eq:gauss3}) is to write its left hand side in
operatorial form. Since the Hamiltonian that dictates the $\tau$-evolution
vanishes, the expression on the right hand side follows.

Thus we arrive to the following expression for $\Gamma(A)$:
\begin{eqnarray}
\Gamma (A) &= & \frac{V}{(2\pi)^2} e {\tilde F}_0 
\int_{0^+}^\infty \frac{dT}{T} e^{- m T} \,
\int_{-\infty}^{\infty} \frac{dp_0}{2 \pi} \nonumber \\
& \times & 
{\rm Tr} \left( e^{- i T \not\hat{p} } e^{- i T \gamma_0 p_0 } \right) \,.
\end{eqnarray}
In terms of the representation $\gamma_0 = \sigma_3$, $\gamma_1=\sigma_1$ and
$\gamma_2=\sigma_2$, we have:
\begin{eqnarray}
\Gamma (A) &=&  \frac{V}{(2\pi)^2} e {\tilde F}_0 
\int_{0^+}^\infty \frac{dT}{T} \, e^{- m T} \,
\int_{-\infty}^{+\infty} \frac{dp_0}{2 \pi} \nonumber\\
&\times& {\rm Tr} [ e^{-i T \sqrt{ 2 e {\tilde F}_0} \hat{\alpha}}
e^{- i T p_0 \sigma_3 } ] \;,
\end{eqnarray}
where
\begin{equation}
{\hat \alpha}\;=\;\left(\begin{array}{cc} 0 & {\hat a} \\ {\hat a}^\dagger & 0
\end{array} \right)
\end{equation}
with ${\hat a} = \frac{{\hat p}_1 - i {\hat p}_2}{\sqrt{ 2 e {\tilde F}_0}}$,
and $[{\hat a},{\hat a }^\dagger ] = 1$. Since the integration over $p_0$ is
over an interval symmetric about $0$, we use the fact that
\begin{equation}
\int_{-\infty}^{+\infty} \frac{d p_0}{2 \pi} \, \sin ( T p_0) \;=\; 0
\end{equation}
to write
\begin{eqnarray}
\Gamma (A) &=&  \frac{V}{2 \pi^2} e {\tilde F}_0 
\int_{0^+}^\infty \frac{dT}{T}  e^{- m T}
\int_0^{+\infty} \frac{dp_0}{2 \pi}  \cos ( T p_0) \nonumber\\
&\times& {\rm Tr}[e^{- i T \sqrt{ 2 e {\tilde F}_0} {\hat \alpha} } ] \;.
\end{eqnarray}
On the other hand, it is simple to show that the operator
\begin{equation}
{\hat \alpha} \;=\; \left(\begin{array}{cc} 0 & {\hat a} \\ {\hat a}^\dagger & 0
\end{array} \right)
\end{equation}
has the eigenvalues $\pm \sqrt{n}$, with $n= \pm1,\pm 2,\ldots$, and no
degeneracy.  Thus
\begin{eqnarray}
\Gamma (A) &=&  \frac{V}{\pi^2} e {\tilde F}_0 
 \int_{0^+}^\infty \frac{dT}{T} \, e^{- m T} \,
\int_0^{+\infty} \frac{dp_0}{2 \pi} \,
\cos ( T p_0) \nonumber\\
&\times& \sum_{n=1}^\infty \cos[ T ( 2 e {\tilde F}_0 n )^{\frac{1}{2}} ]\;.
\end{eqnarray}
We take advantage of the explicit dependence of this result on the
external field to subtract the zero field contribution,
\begin{eqnarray}
{\tilde \Gamma}(A)&=& \frac{V}{\pi^2} e {\tilde F}_0 \int_{0^+}^\infty 
\frac{dT}{T} e^{- m T}
\int_0^{\infty} \frac{dp_0}{2 \pi}  \cos( T p_0) \nonumber\\
&\times& \sum_{n=1}^\infty \left\{ \cos[ T (2 e {\tilde F}_0
n)^{\frac{1}{2}}] - 1 \right\}\;,
\end{eqnarray}
and after performing the $T$ integration,
\begin{eqnarray}
{\tilde \Gamma}(A) &=& - \frac{V}{(2\pi)^2} e {\tilde F}_0
\sum_{n=1}^\infty \, \int_0^{+\infty} \frac{dp_0}{2 \pi} \,
\left\{ \ln[\frac{(p_0 - \kappa_n)^2 + m^2}{p_0^2 + m^2}] \right. \nonumber\\
&+& \left. \ln[\frac{(p_0 + \kappa_n)^2 + m^2}{p_0^2 + m^2}] \right\}\;,
\end{eqnarray}
where we introduced $\kappa_n = \sqrt{ 2 e {\tilde F}_0 n }$.
Finally, the contributions may be rearranged into the expression:
\begin{eqnarray}
{\tilde \Gamma}(A) &=& - \frac{V}{(2\pi)^2} e {\tilde F}_0
\sum_{n=1}^\infty \,
\int_0^{+\infty} \frac{dp_0}{2 \pi} \nonumber\\
&\times& \ln\left[\frac{(p_0^2+m^2+\kappa_n^2)^2 -
4 p_0^2 \kappa_n^2}{(p_0^2 + m^2)^2}] \right]\;.
\end{eqnarray}
The direction of the (constant) Euclidean field ${\tilde F}_\mu$
is arbitrary; we may of course replace ${\tilde F}_0$ everywhere by
$|{\tilde F}|$. The momentum integral is along the `longitudinal' direction,
which not necessarily coincides with the zero (timelike) one. Then:
\begin{eqnarray}
{\tilde \Gamma}(A) &=& - \frac{V}{(2\pi)^2} e {\tilde F}_0 
\sum_{n=1}^\infty \,
\int_0^{+\infty} \frac{dp_\parallel}{2 \pi} \nonumber\\
&\times& \ln\left[\frac{(p_\parallel^2+m^2+2 e|{\tilde F}| n )^2
- 8 p_\parallel^2  e |{\tilde F}| n }{(p_\parallel^2 + m^2)^2} \right]\;,
\end{eqnarray}
which has an explicitly covariant and frame independent form.
The fact that the result depends only on the square of the mass, and
not on its sign, confirms that the parity-odd term vanishes for
this field configuration. In other words, the spectral asymmetry is
zero.

Integrating over the momentum one arrives at a sum of the form 
\begin{equation}
\sum_{n=1}^\infty \, \sqrt{ m + 2 e|{\tilde F}| n  },
\end{equation}
which can be analytically continued to the Hurwitz function
and so it agrees with the result of 
previous calculations \cite{Dittrich}.
The 1+1 dimensional case can be obtained at no cost.

The representation for the Dirac propagator in terms of path integrals
used above is not the only possible one. It is possible to derive the full
fermion propagator by using a set of Grassmannian variables which carry all
the spin information and avoids the presence of the path ordering
operator~\cite{Fradkin,Karanikas:1995ua,Alexandrou,Schmidt}.
This will be discussed separately in detail.

\section{Conclusions}\label{sec:concl}
We have shown that the path integral representation (\ref{eq:fnal}) is,
when expanded in powers of the coupling constant, equivalent to the usual
perturbative series.

Besides, we performed two different kinds of non-perturbative tests:
first, we evaluated the axial anomaly in $1+1$ dimensions and the
Chern-Simons term in $2+1$ dimensions. Amusingly enough, the path
integral is particularly suited for the evaluation of those objects in
a Pauli-Villars regularization scheme, and a subsequent large mass
(cutoff) expansion. Our calculation focuses in the currents, so we can
have perfect control of the gauge invariance of the results.
For both cases, we have seen that the exact results are obtained.

We also considered particle propagation in a constant electromagnetic
field, deriving an expression for the effective action using Migdal's
representation.

This kind of calculations provide, we believe, further support for
the use of these representations in the derivation of Quantum Field
Theory results, either analytically or numerically.
The developments presented here can also be useful for the worldline
in practice given, for instance, the problems with renormalization in
general and the difficulties of the second order formalism with
external fermions and spectral asymmetry originated by the 
Dirac operator such as the Chern-Simons term.

\section*{Acknowledgments}
We thank Holger Gies and Emilio Elizalde for useful comments. 
C.\ D.\ F.\ acknowledges the kind hospitality of the Theoretical
Physics Group at the University of Santiago de Compostela, and
the support of CONICET (Argentina), and  Fundaci{\'o}n Antorchas.
R.A.V. thanks the kind hospitality of the ``Centro At\'omico de
Bariloche''. R.A.V. is supported by
the ``Ram\'on y Cajal'' program. J.S.G. is supported by MCyT and FEDER
with projects FPA2002-01161 and BFM2002-03881 and by the European 
TMR EUCLID HPRN-CT-2002-00325.


\end{document}